\documentclass[11pt]{article}

\usepackage{times}
\usepackage{mathptmx}

\usepackage{graphicx}
\usepackage[fleqn]{amsmath}
\usepackage[square,numbers]{natbib}
\usepackage{fleqn}
\usepackage[letter]{crop}
\usepackage{hyperref}
\usepackage{memhfixc}
\usepackage{amssymb,amsfonts}
\usepackage[margin=2.5cm]{geometry}
\DeclareMathAlphabet{\mathbi}{OML}{cmm}{b}{it} 

\newcommand{\bel}{\begin{equation}\label}
\newcommand{\ee}{\end{equation}}
\newcommand{\beq}{\begin{eqnarray}\label} 
\newcommand{\eeq}{\end{eqnarray}} 
\newcommand{\bc}{\begin{center}} 
\newcommand{\ec}{\end{center}} 
\newcommand{\ben}{\begin{enumerate}}
\newcommand{\een}{\end{enumerate}}
\newcommand{\bit}{\begin{itemize}}
\newcommand{\eit}{\end{itemize}}
\newtheorem{theorem}{Theorem}

\DeclareMathAlphabet{\mathbi}{OML}{cmm}{b}{it} 

\newcommand{\bom}{\mbox{\boldmath$\omega$}}
\newcommand{\bu}{\mathbi{u}}
\newcommand{\bU}{\mathbi{U}}
\newcommand{\bx}{\mathbi{x}}
\newcommand{\rem}[1]{}
\newcommand{\non}{\nonumber}

\newcommand{\bdB}{\mathbi{B}}
\newcommand{\bD}{\mathbi{D}}
\newcommand{\bcapom}{\mathbi{\Omega}}
\newcommand{\bk}{\mbox{\boldmath$\hat{k}$}}
\newcommand{\bS}{\mathbi{S}}

\begin{document}
\title{Stretching and folding processes \\in the 3D Euler and Navier-Stokes equations}
\author{
J. D. Gibbon and D. D. Holm\\
Department of Mathematics
\\Imperial College London
\\London SW7 2AZ, UK}
\date{
UTAM Symposium on Understanding Common Aspects of Extreme Events in Fluids
\\
{\it Procedia IUTAM} {\bf 9} (2013) 25 -- 31
}

\maketitle
\makeatother
\maketitle
\tableofcontents

\begin{abstract}
Stretching and folding dynamics in the incompressible, stratified 3D Euler and Navier-Stokes equations are reviewed in 
the context of the vector $\bdB = \nabla q\times\nabla\theta$ where $q=\bom\cdot\nabla\theta$. The variables $\theta$ 
is the temperature and $\bdB$ satisfies $\partial_{t}\bdB = \mbox{curl}\,(\bu\times\bdB)$. 
These ideas are then discussed in the context of the full compressible Navier-Stokes equations where $q$ 
takes the two forms $q = \bom\cdot\nabla\rho$ and $q = \bom\cdot\nabla(\ln\rho)$.
\end{abstract}

\maketitle

\section{Introduction}\label{intro}

\subsection{Stretching and folding in the incompressible 3D Euler equations}\label{sfeuler}

\noindent
The most fundamental of all equations in three-dimensional fluid dynamics are the incompressible Euler 
equations
\bel{3Deul1a}
\frac{D\bu}{Dt} = -\nabla p\,,\qquad\quad \mbox{div}\,\bu = 0\,,
\ee
which can also be expressed as 
\bel{3Deul1b}
\partial_{t}\bu - \bu\times\bom = -\nabla \big(p + \tfrac12 u^{2}\big)\,.
\ee
$\bu(\bx,\, t)$ is the velocity field and $\bom = \mbox{curl}\,\bu$ the vorticity of the fluid. The material 
derivative is defined by
\bel{matdef}
\frac{D~}{Dt} = \partial_{t} + \bu\cdot\nabla\,.
\ee
Following equation (\ref{3Deul1b}) the vorticity field $\bom$ satisfies
\bel{3Deul2}
\partial_{t}\bom - \mbox{curl}(\bu\times\bom) = 0\,,
\ee
which can also be written in the familiar vortex stretching format
\bel{3Deul3}
\frac{D\bom}{Dt} = \bom\cdot\nabla\bu \equiv \textbf{S}\,\bom\,,
\ee 
where $S_{ij} = \tfrac12\left(u_{i,j}+u_{j,i}\right)$ is the rate of strain matrix.  It is generally 
acknowledged that the stretching and folding processes caused by the rapid alignment or 
anti-alignment of $\bom$ with positive or negative eigenvectors of $\textbf{S}$ roughen 
Euler data very quickly.  In fact, the 3D incompressible Euler equations have an array of very 
weak solutions \cite{Sh97,Brenier99,BT07,deLSz09,deLSz10,BT10,Wied11,BTW12}, but the 
Leray-type weak solutions associated with the Navier-Stokes equations are unknown \cite{ConstAMS}.
Our lack of knowledge forces us to make some assumptions about the existence of solutions 
of both the incompressible Euler and Navier-Stokes equations in order to perform formal 
manipulations \cite{BKM,MB01,Aussois}. Likewise in later sections we also discuss the 
compressible case in the same spirit. This paper aims to marry the ideas on stretching and 
folding processes in incompressible flows developed by the authors in \cite{JDGD2H10,JDGD2H11} with 
their work on compressible flows \cite{JDGPRE12}. Here, two forms of the projection of the vorticity 
$\bom$ onto the gradient of the mass density $\rho(\bx,\,t)$ are discussed in \S\ref{PRE}. 

The familiar vortex stretching format expressed in (\ref{3Deul3}) also appears in a different context. 
Let $\bu$ be a divergence-free Euler flow and let $\theta$ and $q$ be two arbitrary (for now) passive 
scalars riding on this flow
\bel{qtheta}
\frac{D\theta}{Dt} = 0\qquad\qquad \frac{Dq}{Dt} = 0\,.
\ee
Then it has been shown that the vector \cite{KurgPis00,KurgTat87,Kurg02} (see also \cite{JDGD2H10,JDGD2H11})
\bel{Bdef}
\bdB = \nabla q\times\nabla\theta
\ee
satisfies
\bel{B1}
\partial_{t}\bdB - \mbox{curl}\,(\bu\times\bdB) = 0,\,\qquad\Longrightarrow\qquad
\frac{D\bdB}{Dt} = \bdB\cdot\nabla\bu\,,
\ee
which is exactly (\ref{3Deul3}). Thus the $\bdB$-field shares the same stretching and folding properties of the 
$\bom$-field, in much the same manner as a magnetic field in ideal MHD \cite{HKM1,HKM2}.
\par\smallskip
The context of this result can be seen by considering the Euler equations with constant rotation $2\bcapom$ 
and buoyancy (proportional to the temperature $\theta$), written in the dimensionless form 
\bel{ens8a}
\frac{D\bu}{Dt} + 2\bcapom\times\bu +  a_{0}\theta\,\bk = - \nabla p\,,
\qquad\qquad \frac{D\theta}{Dt} = 0\,,
\ee
with $a_{0}$ as a dimensionless constant. 
The equation for the vorticity $\bom_{rot} = \bom + 2\bcapom$ is now 
\bel{omrot1}
\frac{D\bom_{rot}}{Dt} + a_{0}\nabla^{\perp}\theta = \bom_{rot}\cdot\nabla\bu
\ee
where $\nabla^{\perp} = (\partial_{y},\,-\partial_{x},\,0)$.
With $q$ taken as a potential vorticity $q=\bom_{rot}\cdot\nabla\theta$ it is easily seen that
\beq{q1A}
\frac{Dq}{Dt} = \left(\frac{D\bom_{rot}}{Dt} - \bom_{rot}\cdot\nabla\bu \right)\cdot\nabla \theta 
+ \bom_{rot}\cdot\nabla\left(\frac{D\theta}{Dt}\right)\,,
\eeq
which is a geometric relation reminiscent of Ertel's Theorem \cite{Ertel42}. Because 
$\nabla^{\perp}\theta\cdot\nabla\theta = 0$ it is clear that 
\bel{q1B}
\frac{Dq}{Dt} = 0
\ee
and so (\ref{qtheta}) is satisfied. It can now be seen that the stretching and folding properties of $\bdB =  
\nabla q\times\nabla\theta$ in equation (\ref{B1}) are now lifted to higher gradients of $\bom$ because 
$\bdB$ contains a gradient of $\bom$ in projection and two gradients of $\theta$.

\subsection{Stretching and folding in the incompressible 3D Navier-Stokes equations}\label{sfns}

The stratified Navier-Stokes equations are the viscous equivalent of (\ref{ens8a}) 
\bel{NSens8a}
\frac{D\bu}{Dt} + a_{0}\theta\,\bk = Re^{-1}\Delta\bu - \nabla p\,,
\ee
\bel{NSens8b}
\frac{D\theta}{Dt} = \big(\sigma Re\big)^{-1}\Delta\theta\,.
\ee
Here, the potential vorticity $q = \bom\cdot\nabla\theta$ is no longer a material constant (the `rot' suffix has been 
dropped) but, instead, evolves according to
\beq{ens9}
\frac{Dq}{Dt} &=& \left(\frac{D\bom}{Dt} - \bom\cdot\nabla\bu \right)\cdot\nabla \theta
+ \bom\cdot\nabla\left(\frac{D\theta}{Dt}\right)\non\\
&=& \big(Re^{-1}\Delta\bom - \nabla^{\perp}\theta \big)\cdot\nabla \theta
+ \bom\cdot\nabla\left[(\sigma Re)^{-1}\Delta\theta\right]
\non\\
&=& \mbox{div}\big\{Re^{-1}\Delta\bu\times\nabla \theta 
+ (\sigma Re)^{-1}\bom\,\Delta\theta \big\}\,.
\eeq
The material advection property no longer holds but the introduction of a \textit{pseudo-velocity field} 
$\bU_{q}$ transforms (\ref{ens9}) into a continuity equation
\bel{ens10}
\partial_t q +  \mbox{div}\,(q\,\bU_{q}) = 0\,,
\ee
thus making $q$ a potential vorticity density, and where $\bU_{q}$ is defined through
\bel{ens11}
q\big(\bU_{q} - \bu\big) = - Re^{-1}\big(\Delta\bu\times\nabla \theta  + 
\sigma^{-1}\bom\Delta\theta\big)\,.
\ee
Moreover, $\theta$ evolves according to
\beq{ens12}
\partial_{t}\theta + \bU_{q}\cdot\nabla\theta &=& \partial_{t}\theta + \bu\cdot\nabla\theta - 
Re^{-1}q^{-1}\left\{\Delta\bu\times\nabla \theta + \sigma^{-1}\bom\Delta\theta\right\}
\cdot\nabla\theta\non\\
&=& \partial_{t}\theta + \bu\cdot\nabla\theta - \big(\sigma Re\big)^{-1}\Delta\theta\non\\
&=& 0\,.
\eeq
The formal result for the stratified Navier-Stokes equation is\,:
\begin{theorem}\label{Bthm}
The scalar quantities $q$ and $\theta$ satisfy
\bel{ens13}
\partial_{t} q + \mbox{div}\,\big(q\,\bU_{q}\big) = 0\,,\qquad \qquad
\partial_{t}\theta + \bU_{q}\cdot\nabla\theta = 0\,,
\ee
and $\bdB = \nabla q\times\nabla \theta$ satisfies the stretching and folding relation
\bel{ens14}
\partial_{t}\bdB - \mbox{curl}\,(\bU_{q}\times\bdB) = \bD_{q}\,,
\ee
where the divergence-less vector $\bD_{q}$ is given by 
\bel{ens15}
\bD_{q} = - \nabla(q\,\mbox{div}\,\bU_{q})\times\nabla\theta\,,
\ee
and the pseudo-velocity $\bU_{q}$ is defined as in (\ref{ens11}). Moreover, for any 
surface $\bS(\bU_{q})$ moving with the flow $\bU_{q}$
\bel{ens16}
\frac{d}{dt}\int_{\bS(\tiny\bU_{q})} \bdB \cdot d\bS = \int_{\bS(\tiny\bU_{q})}\bD_{q}\cdot d\bS\,.
\ee
\end{theorem}
The introduction of a pseudo-velocity field $\bU_{q}$ that `hides' the dissipation is based originally on an idea due 
to Haynes \& McIntyre in an atmospheric context \cite{HMc87,HMc90}. The two obvious drawbacks are that firstly the 
divergence-free property is lost ($\mbox{div}\,\bU_{q} = O\left( Re\right)^{-1}$) and secondly the transformation 
(\ref{ens11}) fails at zeros of $q$ where a change of topology of vortex lines could occur. Moreover, the stretching and 
folding relation (\ref{ens14}) now has a non-zero right hand side $\bD_{q}$ defined in (\ref{ens15}) that drives and 
modifies the process. Numerical studies on reconnection (Herring, Kerr \& Rotunno \cite{HKR94}) suggest that in the 
early or intermediate stages of a flow this divergence may be small because.

\section{The compressible Navier-Stokes equations}\label{PRE}

The aim of this section is to now develop the ideas of \S\ref{intro} in the context of compressible flows 
\cite{JDGPRE12}. In this context the mass density $\rho$ plays the role of $\theta$ even though it is 
not a passive quantity. The 3D compressible Navier-Stokes equations are expressed 
as \cite{LLhydro}
\beq{comp1}
\rho\,\frac{D\bu}{Dt}  =  \mu\Delta\bu\ - \nabla \varpi 
\qquad\qquad
\varpi = p - (\mu/3+\mu^v)\mbox{div}\,\bu
\eeq
where $\rho$ and the temperature $\theta$ satisfy
\bel{comp2}
\frac{D\rho}{Dt} + \rho\,\mbox{div}\,\bu = 0
\qquad\qquad
c_{v}\frac{D\theta}{Dt} = \frac{p}{\rho}\,\mbox{div}\,\bu + Q 
\ee
$\mu$ is the shear viscosity and $\mu^v$ is the volume viscosity, both of which are  taken as constitutive constants of the fluid. 
An ideal gas equation of state $p = R \rho \theta$ has been chosen to relate pressure, $p$, temperature, $\theta$, and mass 
density, $\rho$. In addition, $R$ is the gas constant,  $c_{v}$ the specific heat is constant and $Q$ is the heating rate, which 
is assumed to be known.  However, the geometric considerations that follow are universal for any viscous compressible fluid flow 
because the dynamics of the temperature and the choice of equation of state will not affect our considerations of the transport 
dynamics of the projection $q$ in (\ref{qr1}). It is well known that compressible flows have weak shock solutions which have 
subtle properties \cite{DH1,DH2,DH3}. In the following our manipulations are based on the assumption that the necessary 
differentiations are allowed although there well may be some situations when they are not. 

\subsection{1st projection}\label{proj1}

The first choice for $q$ as a projection of $\bom$ onto $\nabla\rho$ is
\bel{qr1}
q = \bom\cdot\nabla\rho\,.
\ee
According to equations (\ref{comp1}) and (\ref{comp2}) the vorticity $\bom$ evolves according to
\bel{Dom}
\partial_{t}\bom - \mbox{curl}\,(\bu\times\bom) = \mu\rho^{-1}\Delta\bom + 
\nabla\left(\rho^{-1}\right)\times\left[\mu\Delta\bu - \nabla \left(\varpi + \tfrac12 u^{2}\right)\right]
\ee
Using (\ref{q1A}) with $\theta$ replaced by $\rho$, $q$ now satisfies 
\bel{q-calc2}
\partial_{t}q + \mbox{div}\,q\bu + \mbox{div}\big(\bom\rho\,\mbox{div}\,\bu
- \mu\Delta\bu\times\nabla(\ln\rho)\big) = 0\,.
\ee
Note that the pressure terms have disappeared with no approximation. Now we apply an observation 
of Haynes and McIntyre \cite{HMc87,HMc90} that first arose in atmospheric physics and which leads 
to the definition of a current density $\mathbf{J}$
\bel{J1def}
\mathbf{J} = q\bu + \bom\rho\,\mbox{div}\,\bu - \mu\Delta\bu\times\nabla(\ln\rho)
+ \nabla\phi\times \nabla \psi(\rho)\,,
\ee
where $\phi$ is an undetermined gauge potential and $\psi$ is an arbitrary differentiable function of $\rho$. 
\par\smallskip
(\ref{q-calc2}) and (\ref{proj2B}) can be rewritten in the \emph{quasi-conservative} form
\bel{q1d}
\partial_{t}q + \mbox{div}\,\mathbf{J} = 0
\quad\hbox{and}\quad
q\,\partial_{t}\rho + \mathbf{J}\cdot\nabla\rho = 0\,.
\ee
The relation $\mathbf{J}\cdot\nabla\rho=q\,\mbox{div}\,\rho\bu $ allows zero projection, $q=0$, by the second equation in (\ref{q1d}). 
Thus, the projection $q$ may vanish anywhere in the flow, but it cannot be maintained, because $\mbox{div}\,\mathbf{J} \ne0$. Together, 
the equations in (\ref{q1d}) imply a family of conserved quantities, since
\bel{cons}
\partial_{t} (q\Phi'(\rho)) +  \mbox{div}\,(\mathbf{J}\Phi'(\rho))=0
\,,\ee
for any function $\Phi'(\rho)=d\Phi/d\rho$.
The  conserved  densities $q\Phi'(\rho)={\rm div}(\Phi(\rho)\bom)$ in (\ref{cons}) possess quite  different flow properties from those of mass, 
energy  and momentum. 

\subsection{2nd projection}

There is a second choice for $q$ as a projection of $\bom$ onto $\nabla\rho$ is
\bel{qr2}
q = \bom\cdot\nabla(\ln\rho)\,,
\ee
in which case
\bel{proj2B}
\frac{Dq}{Dt} = \left(-\bom\,\mbox{div}\,\bu +\mu \rho^{-1}\Delta\bom\right)\cdot\nabla\left(\rho^{-1}\right) 
-\bom\cdot\nabla\left(\mbox{div}\,\bu\right)\,.
\ee
Equivalent to (\ref{q-calc2}) we have 
\bel{proj2B}
\partial_{t}q + \mbox{div}\,(q\bu) + \mbox{div}\,
\left\{\mu\Delta\bu\times\nabla\left(\rho^{-1}\right) + \bom\,\mbox{div}\,\bu\right\} = 0\,.
\ee
Therefore a second definition of the current density $\mathbf{J}$ is 
\bel{J2def}
\mathbf{J} = q\bu + \bom\,\mbox{div}\,\bu + \mu\Delta\bu\times\nabla\left(\rho^{-1}\right)
+ \nabla\phi\times \nabla \psi(\rho)\,,
\ee
and similar conclusions to those in (\ref{q1d}) and (\ref{cons}) follow.

\subsection{Physical interpretation?}

Equations (\ref{q1d}) are purely kinematic, because the projection taken against $\nabla\rho$ has removed any dependence on the temperature 
and the choice of equation of state. Nonetheless, equations 
(\ref{q1d}) have been derived without approximation from the Navier-Stokes  fluid equations for compressible motion and mass transport. 
Moreover, their analogues also occur and have been found useful in other areas  of  fluid dynamics, particularly in the dynamics of the ocean 
and atmosphere  \cite{HMR85,HMc87}. There (\ref{q1d}) was locally interpreted as a type of \emph{impermeability theorem} for the 
\emph{quasi-Lagrangian} transport of the projection $q$ and the mass density $\rho$ by a pseudo-velocity $\mathbf{U}_{q}$ defined by
\bel{Udef}
q\bU_{q}=\mathbf{J}
\ee
Namely, the projections within $q= \bom\cdot\nabla\rho$ and $q= \bom\cdot\nabla(\ln\rho)$ cannot be transported\footnote{By the definition of $\mathbf{J}$ the pseudo-velocity $\bU_{q}=\mathbf{J}/q$ is determined only up to the addition of the curl of a vector proportional to $\nabla\rho$.} 
by the pseudo-velocity $\bU_{q}$ across level sets of the mass density, $\rho$. These equations may also be useful in the study of stretching and 
folding in compressible fluid flows, as investigated in the atmospheric context in \cite{JDGD2H10,JDGD2H11}.

\section{Conclusion}

Although the usual concerns of compressible flows focus on shock formation \cite{DH1,DH2,DH3}, some final remarks are in order about how the gradient 
of the projection $q$ participates in stretching and folding in compressible Navier-Stokes fluid flows as in Theroem \ref{Bthm}. In preparation, 
we rewrite equations (\ref{comp2}) and (\ref{q1d}) in terms of the pseudo-velocity $\bU_{q}$ defined in (\ref{Udef}) as
\bel{q1d2}
\partial_{t}q + \mbox{div}\,\left(q\,\bU_{q}\right) = 0
\qquad\qquad
\partial_{t}\rho + \bU_{q}\cdot\nabla\rho = 0
.\ee
Note however that  $\mbox{div}\,\bU_{q} \ne 0$. With the definition 
\bel{Bdef}
\bdB = \nabla q\times\nabla \rho
\ee
a direct computation shows that
$\bdB$ satisfies
\bel{ceeqn1}
\partial_{t}\bdB - \mbox{curl}\,(\bU_{q}\times\bdB) = \bD_{q}
\ee
where 
\bel{eeqn2}
\bD_{q} = - \nabla\left(q\,\mbox{div}\,\bU_{q}\right)\times\nabla\rho
\ee
as in Theorem \ref{Bthm}. 
\par\smallskip
Based on the pseudo-velocity defined in (\ref{Udef}), the left hand side of (\ref{ceeqn1}) makes it clear that the vector $\bdB$ undergoes the same type of stretching and folding processes driven by the $\bD$-vector on the right hand side as occurs in the vorticity equation (\ref{Dom}). This is remarkable, because $\bdB = \nabla q\times\nabla\rho$ contains information not only about $\nabla\bom$ but also about $\nabla\rho$ and even $\nabla^{2}\rho$. Thus, the stretching and folding in the original vorticity equation (\ref{Dom}) compounds itself in the same form in the $\bdB$-equation (\ref{ceeqn1}), but with higher spatial derivatives.



\end{document}